\documentclass{PoS}

\title{The Advanced X-ray Timing Array (AXTAR): \\ A US MIDEX Mission Concept}

\ShortTitle{AXTAR}

\author{\speaker{Paul S. Ray}, Bernard F. Phlips, Kent S. Wood\\
        Naval Research Laboratory, Washington, DC, 20375, USA\\
        E-mail: \email{paul.ray@nrl.navy.mil}}

\author{Deepto Chakrabarty, Ronald A. Remillard\\
Massachusetts Institute of Technology, Cambridge, MA, 02139, USA \\
        E-mail: \email{deepto@space.mit.edu}}

\author{Colleen A. Wilson-Hodge\\
	NASA/Marshall Space Flight Center, Huntsville, AL, 35812, USA \\
	E-mail: \email{colleen.wilson@nasa.gov}}
	
\abstract{AXTAR is a NASA MIDEX mission concept for X-ray timing of
  compact objects that combines very large collecting area, broadband
  spectral coverage, high time resolution, highly flexible scheduling,
  and an ability to respond promptly to time-critical targets of
  opportunity. It is optimized for submillisecond timing of bright
  Galactic X-ray sources in order to study phenomena at the natural
  time scales of neutron star surfaces and black hole event horizons,
  thus probing the physics of ultradense matter, strongly curved
  spacetimes, and intense magnetic fields. AXTAR's main instrument is
  a collimated, thick Si pixel detector with 2--50 keV coverage and
  over 3 square meters effective area. For timing observations of
  accreting neutron stars and black holes, AXTAR provides at least a
  factor of five improvement in sensitivity over the RXTE PCA. AXTAR
  also carries a sensitive sky monitor that acts as a trigger for
  pointed observations of X-ray transients in addition to providing
  high duty cycle monitoring of the X-ray sky. We review the science
  goals and design choices that face a next generation timing mission. 
  We then describe the technical concept for AXTAR and summarize a preliminary mission design study at the NASA/MSFC Advanced
  Concepts Office.}

\FullConference{Fast X-ray timing and spectroscopy at extreme count
  rates: Science with the HTRS on the International X-ray Observatory
  - HTRS2011,\\ February 7--11, 2011\\ Champ\'{e}ry, Switzerland}

\begin{document}

\section{Introduction}

The properties of ultradense matter and strongly curved spacetime and
the behavior of matter in the extreme environments near compact
objects are among the most fundamental problems in astrophysics. X-ray
timing measurements have powerful advantages for studying these
problems \cite{lamb04}.  The X-ray band contains most of the power
emitted by accreting neutron stars and black holes, and this radiation
is relatively penetrating even in these complex environments.  The
millisecond X-ray variability of these objects encodes their basic
physical parameters, and intepretation of this variability is
relatively straightforward for rotating or orbital origins.  In many
cases, the properties of the X-ray variability allow extremely precise
measurements and detailed quantitative inferences.

The scientific promise of X-ray timing has been spectacularly
demonstrated by the success of NASA's Rossi X-ray Timing Explorer
(RXTE; effective area $A_{\rm eff}=0.6$~m$^2$; launched 1995), which
has revealed an extraordinary range of previously unknown variability
phenomena from neutron stars and black holes.  However, redeeming that
promise to exploit these phenomena as tools for answering fundamental
astrophysical questions will require a more sensitive follow-on
mission.  A detailed scientific case for such a mission was first
explored at the conference {\em X-Ray Timing 2003; Rossi and Beyond}
in Cambridge, MA, USA \cite{rossi03}, and has been discussed in more
detail at this 2010 meeting in Champ\'{e}ry, Switzerland.  Many of the
issues concerning the fundamental properties of neutron stars and
black holes have been identified as high priority scientific questions
by the 2010 U.S. Decadal Survey of Astronomy and Astrophysics
\cite{NWNH}.

In this paper we describe the Advanced X-ray Timing Array (AXTAR), a
new mission concept with significantly larger effective area than
RXTE. AXTAR was originally proposed as an ~$\sim$8~m$^2$ medium-class
probe concept in the 2007 NASA Astrophysics Strategic Mission Concepty
Study call \cite{c+08}.  More recently, we have been
developing as a $\sim$4~m$^2$ Medium Explorer (MIDEX) class mission
concept \cite{Ray2010}.  

\section{Design Choices}

The first, critical design decision for any future X-ray timing
mission is whether to include a focusing optic or to adopt a
collimated detector as was done with RXTE.  Each architecture has
advantages and disadvantages and is particularly well matched to
particular science questions.  

{\bf Focusing optics.} The primary benefit of a focusing optic is that
the X-rays from the source of interest are focused onto a small
detector region.  This allows a large reduction in the background
counting rate and enables studies of faint sources (e.g.,
rotation-powered pulsars, AGN, iron line sources, etc.).  In addition,
the small detector area can reduce the required power and make it
easier to achieve very good (e.g. $< 200$ eV) energy resolution. The
drawbacks of focusing systems include the fact that it is difficult to
achieve good efficiency for higher energy X-rays, since mirror systems
become particularly challenging above 10~keV. Such coverage can be
critical to many science topics in X-ray timing, particularly with
respect to X-ray binaries.  To get significant effective area at high
energies, one is driven to designs with small grazing angles and long
focal lengths.  As a result, many focusing X-ray telescopes have large
masses, which increases cost, and large moments of inertia, which
makes flexible scheduling and rapid repointing difficult.  An additional
challenge with focusing optics is that the concentration of flux onto
a small detector area can lead to significant deadtime effects,
limiting the ability to observe bright targets.

One can also choose the focusing approach for opportunistic or
serendipitous reasons, such as the community plans for the flagship
focusing X-ray astronomy mission IXO\footnote{Now renamed ATHENA and
  under redesign as of early 2011}.  The HTRS \cite{htrs} instrument
(a major topic of this conference) would add powerful high-count-rate
timing capabilities to IXO/ATHENA.  It provides 1--2 m$^2$ collecting
area over the band 0.3 to 10 keV, $\mu$s time resolution and minimal
deadtime on bright sources, mitigating one of the above drawbacks.
However, since X-ray timing was not a primary requirements driver for
this mission, some of the other drawbacks (hard X-ray response, rapid
repointing) remain.  Another approach to mitigating some drawbacks is
taken by the proposed NICER experiment \cite{nicer}, which adopts
single-bounce X-ray concentrators instead of multi-mirror true imaging
systems. This provides good background rejection with increased area
efficiency and reduced mass. In addition, they use an array of many
small optics, which enables them to use a short focal length of only
1.5 meters for a more compact, agile design.

{\bf Collimated detectors.} The alternative technical approach, a
collimated design, has a different set of strengths and
weaknesses. Previous timing missions (e.g, SAS-3, EXOSAT, Ginga, RXTE,
and the forthcoming Indian ASTROSAT) employed collimated proportional
counters, but these are too heavy for significantly larger effective
areas. However, one can instead substitute silicon detectors and
achieve a substantial reduction in mass.  In that case, the main
strength of collimation is that, for the same cost and mass, one 
can achieve significantly larger effective areas than with focusing.
Also, a collimated design requires no optics, designs can much more easily
accommodate high energies.  Morever, without long optical benches, the
moment of inertia can be small, allowing rapid repointing.  Since the
source photons are not concentrated on the detectors, achieving low
deadtime on bright sources is straightforward.  

The primary drawback of collimated designs is that the lack of
focusing means that the instrumental and diffuse X-ray background
rates will be considerably higher. For bright sources, this is not a
problem, but faint source observations, particularly those that depend
on accurate knowledge of the background rate, will be impaired. In
addition, the need to instrument a very large area of detectors
implies that the power available for the detector readout will be
limited. This makes achieving very good energy resolution more
difficult and power can become a limiting factor.

{\bf Science requirements.} 
Ultimately, the choice of focusing versus collimation is driven by
one's science requirements.  Our primary scientific objectives require
observatons of accreting neutron stars and black holes in Galactic
X-ray binaries.  These are bright sources which can also have intense
X-ray flares or bursts, leading to high count rates.  Many of the
interesting timing phenomena discovered by RXTE are strongest in the
hard X-ray band, requiring sensitivity above 10~keV.  Finally, many of
these sources transition between different accretion/spectral states,
and some key timing phenomena are preferentially present in a
particular state; thus, flexible scheduling and rapid repointing
ability are required.  These considerations lead us to a collimated
design for the AXTAR concept presented below.  The proposed LOFT
mission \cite{loft}, which is driven by many of the same
considerations, has also adopted a collimated approach. 

\section{Instrument Description}

AXTAR hosts two science instruments, the Large Area Timing Array
(LATA), and the Sky Monitor (SM). Both are based on large-area ($10
\times 10$ cm) 2-mm thick silicon pixel detectors, which have been
developed at NRL. The thick detectors enable good efficiency up to at
least 50 keV. The detectors are divided into $2.5 \times 2.5$ mm
pixels.  On the LATA, the pixilation keeps the capacitance low,
enabling 600 eV energy resolution with a low power readout ASIC, as
well as ensuring that dead time is not an issue. For the SM, the 2-D
position resolution of the detectors is exploited to form the basis
for a 2-D wide-field coded aperture mask camera with a $40^\circ\times
40^\circ$ field of view.  High duty-cycle coverage of a large fraction
of the sky is achieved by mounting several of these cameras on the
spacecraft.

\begin{figure}
\begin{center}
\includegraphics[width=2.5in]{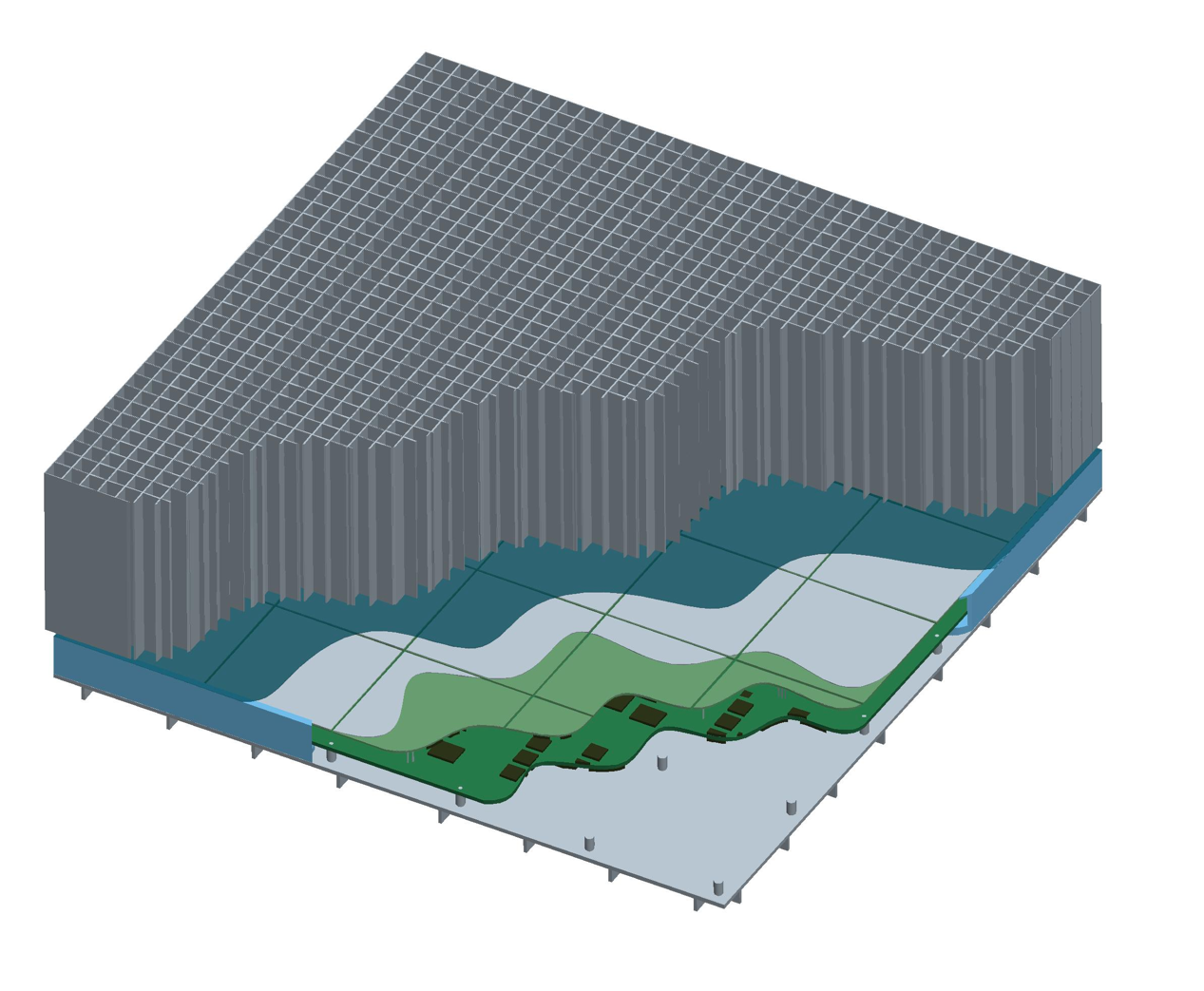}
\hspace{0.5in}
\includegraphics[width=2.5in]{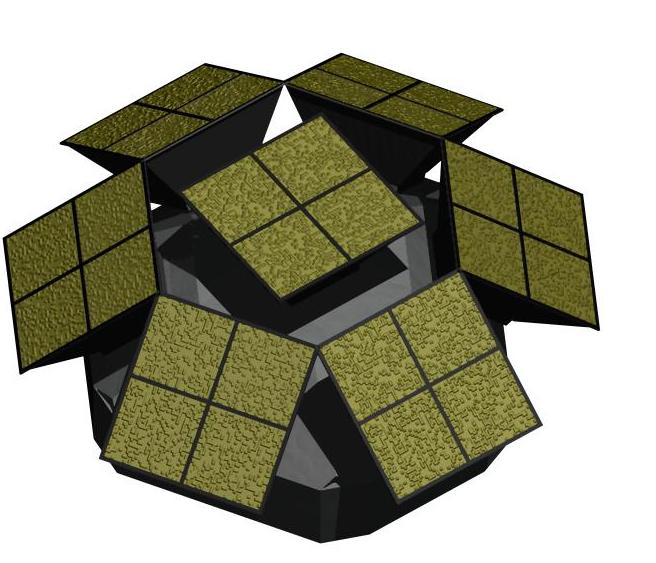}
\end{center}
\caption{\textbf{Left:} Cutaway rendering of a LATA supermodule
  consisting of a 5 $\times$ 5 array of 10 $\times$ 10 cm
  detectors. The components from top to bottom are the collimator,
  light shield, silicon detectors, interposer board, and digital
  board, mounted in a box that provides support and shielding. Note
  that the collimator cell size is not to scale. \textbf{Right:} A
  cluster of 7 SM cameras. \label{fig:supermodule}}
\end{figure}

The overall performance parameters are shown in Table 1 and the instruments are described in more detail in \cite{Ray2010}.

\begin{table}[t]
\caption{Mission Requirements}
\scriptsize
\begin{tabular}{p{1.0in}p{0.74in}p{2.0in}p{1.5in}}
\hline\hline
Parameter &	Baseline & Drivers & Technology Factors \\
\hline\hline
    \multicolumn{4}{c}{\em Large Area Timing Array (LATA)}\\
\hline
Effective Area & 3.2 m$^2$& NS radius, BH QPOs  & Mass, cost, power \\
Minimum Energy & 1.8 keV	& Source states, absorption meas., soft srcs	& Detector electronics noise \\  
Maximum Energy & $>$30 keV	& BH QPOs, NS kHz QPOs, Cycl. lines & Silicon thickness \\ 
Deadtime	& 10\%@10~Crab$^*$ & Bright sources, X-ray bursts & Digital elec.\,design, pixel size  \\ 
Time Resolution & 1 $\mu$s & 	Resolve ms oscillations & Shaping time, GPS, Digital elec. \\
\hline
	\multicolumn{4}{c}{\em Sky Monitor (SM)} \\
\hline
Sensitivity (1~d) &	$<5$ mCrab$^*$ &	Faint transients, multi-source monitoring	& Camera size/weight/power\\ 
Sky Coverage &	$>2$ sr & 	TOO triggering, multi-source monitoring &	\# cameras vs. gimbaled designs\\ 
Source Location & 1 arcmin & Transient followup & Pixel size, camera dimensions \\
\hline
  \multicolumn{4}{c}{\em AXTAR Mission}\\
\hline
Solar Avoidance Ang & 	30$^\circ$ &	Access to transients	& Thermal/Power design \\ 
Telemetry Rate & 1 Mbps &	Bright sources & Ground stations/TDRSS costs\\ 
Slew Rate & $>6^\circ$\,min$^{-1}$& Flexible scheduling, fast TOO response & Reaction wheels\\
\hline
\multicolumn{4}{l}{$^*$1 Crab = $3.2\times 10^{-8}$ erg~cm$^{-2}$~s$^{-1}$ (2--30 keV)}
\end{tabular}
\end{table}

\section{Mission Concept Study}

In this section, we briefly summarize the baseline design resulting
from a mission concept study at the MSFC Advanced Concepts Office. For
purposes of this study, we hypothesize a 2014 call for proposals for a
$\sim$\$300M (excluding launch) MIDEX class mission to be launched in
2019. Full details of the study input parameters and results are
described in \cite{Ray2010}.

\begin{figure}
\begin{center}
\includegraphics[width=4.25in]{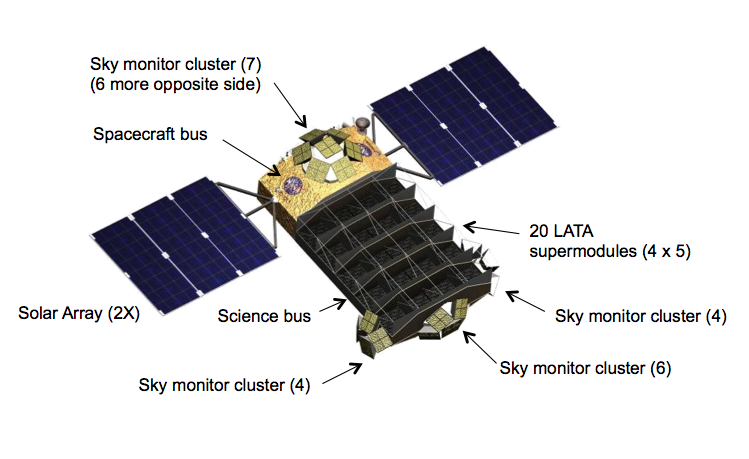}
\end{center}
\caption{AXTAR spacecraft configuration with 20 LATA
  supermodules. This configuration is within the volume and payload
  mass limits for a Taurus~II launcher, and will also easily work with
  a Falcon 9 launcher. \label{fig:TaurusII_config}}
\end{figure}

The optimal orbit was determined to be 585 km altitude circular orbit
with as low an inclination as possible. Given an initial mass estimate
of 2000 kg, two launch vehicle candidates were selected: the Orbital
Sciences Corporation's Taurus II and the SpaceX Falcon 9. The size of
the Taurus II determined the spacecraft configuration and limited the
science instruments to 20 LATA supermodules and 27 sky monitor
cameras, for a total gross mass (dry mass, inert mass, and propellant)
of 2650 kg (including 30\% contingency) and a total power budget of
1583W (including spacecraft systems, science instruments, and a 30\%
growth margin).  Spacecraft structures consisted of 2020-T351 aluminum
panels, struts, and frames for component mounting which also double as
radiators for thermal management. Cosmic and solar radiation shielding
is included in the spacecraft structural mass. The communications
system consists of an S-band transmitter for spacecraft telemetry and
communications and an X-band transmitter for science data downloads,
using two ground stations (Southpoint Hawaii and Kourou Guiana),
allowing expected continuous data rates from the LATA and Sky Monitor
with headroom for over 6 LATA observations per day of very bright (several Crab) sources. 

The avionics system consists of Proton 200 flight computers (TRL 6)
and Surrey data recorders (TRL 8).  Attitude knowledge is provided by
TRL 8 star trackers and IMUs. AXTAR's modest slewing and pointing
requirements, 180 deg in 30 minutes and $< 5$ arcmin, respectively,
allow use of off-the-shelf reaction wheels with magnetic torque rods
for contingency and angular momentum dumping. Inertial pointings of up
to 28 hours are allowed. Thermal control is achieved using passive
components including multilayer insulation, high emissivity paint,
coatings, and heaters, to maintain acceptable temperature ranges. To
allow the spacecraft to be de-orbited at the end of the mission, a
propulsion system was included. The mission concept study found that
AXTAR was straightforward from an engineering point of view, requiring
no new technologies to implement the mission.

\section{Current Status and Plans}

The AXTAR concept continues to be studied in preparation for the next
NASA MIDEX Announcement of Opportunity, which could come as early as
2014. We are pursuing several lines of technical development as well
as studying design alternatives.

Our concept study made clear that a large collimator, such as the one
used on the RXTE PCA, becomes the dominant mass driver for the
instrument when the heavy gas containment vessels are replaced by
lightweight solid state detectors as planned for AXTAR. Therefore,
there is a major mass savings to be had by looking at alternatives.
One option is lead-glass micro-capillary plate collimators, as
currently planned for the LOFT mission.  These can be thin and light,
but their performance is poor at high energies (30 keV and above) and 
achieving a high open fraction ($> 70$\% is challenging).  We
are developing 2-mm thick micromachined tantalum collimators that
could provide excellent high energy performance with a significant
mass reduction that would reduce the expected cost of the AXTAR
mission.

This work was partially supported by the NASA APRA program and
NRL/ONR, as well by internal funding from NASA/MSFC and the MIT Kavli
Institute for Astrophysics and Space Research.


\begin{thebibliography}{99}
\bibitem{lamb04}
Lamb, F.~K. 2004, in {\em X-ray Timing 2003: Rossi and Beyond}, (AIP
Conf.Proc. 714), 3
\bibitem{rossi03}
Kaaret, P., Lamb, F.~K., \& Swank, J.~H. 2004, {\em X-ray Timing 2003:
  Rossi and Beyond} (AIP Conf.Proc. 714) 
\bibitem{NWNH}
Blandford, R.D. et al. 2011, {\em New Worlds, New Horizons in Astronomy and 
  Astrophysics}, (U.S. National Academy Press)
\bibitem{c+08}
Chakrabarty, D. et al. 2008, in {\em A Decade of Accreting Millisecond
  X-ray Pulsars} (AIP Conf.Proc. 1068), 227
\bibitem{Ray2010}
Ray, P.S. et al. 2010, Proc. SPIE, 7732, 773248 
\bibitem{htrs}
Barret, D. et al. 2010, Proc. SPIE, 7732, 77321M
\bibitem{nicer}
Gendreau, K. \& Arzoumanian, Z. 2009, in {\em Astrophysics with All-Sky
  X-Ray Observations --- 3rd International MAXI Workshop}, (RIKEN), 402  
\bibitem{loft}
Feroci, M. et al. 2010, Proc. SPIE, 7732, 77321V
\end{thebibliography}
\end{document}